\documentclass[%
 aip,
rsi,%
 amsmath,amssymb,
 reprint,%
]{revtex4-1}

\usepackage[dvipsnames]{xcolor}
\usepackage{graphicx}
\usepackage{dcolumn}
\usepackage{bm}

\usepackage{dcolumn}
\usepackage{epstopdf}
\usepackage{xcolor}
\usepackage{natbib}
\usepackage{subcaption} 
\usepackage{float}
\usepackage{tabularx}
\usepackage{mathrsfs}
\usepackage{upgreek}
\usepackage[a]{esvect}
\setcitestyle{square}
\usepackage[toc,page]{appendix}
\usepackage[font={small,normalfont}]{caption}
\usepackage[colorlinks=true, allcolors=blue]{hyperref}
\usepackage[normalem]{ulem}

\begin{document}

\title{Electron confinement by laser-driven azimuthal magnetic fields\\during direct laser acceleration}

\author{T. Wang}
\affiliation{Department of Mechanical and Aerospace Engineering, University of California at San Diego, La Jolla, CA 92093}

\affiliation{Center for Energy Research, University of California at San Diego, La Jolla, CA 92093}

\author{Z. Gong}
\affiliation{Center for High Energy Density Science, The University of Texas, Austin, TX 78712}
\affiliation{SKLNPT, KLHEDP, and CAPT, School of Physics, Peking University, Beijing 100871, China}

\author{A. Arefiev}
\email[]{aarefiev@eng.ucsd.edu }

\affiliation{Department of Mechanical and Aerospace Engineering, University of California at San Diego, La Jolla, CA 92093}

\affiliation{Center for Energy Research, University of California at San Diego, La Jolla, CA 92093}

\date{\today}

\begin{abstract}
A laser-driven azimuthal plasma magnetic field is known to facilitate electron energy gain from the irradiating laser pulse. The enhancement is due to changes in the orientation between the laser electric field and electron velocity caused by magnetic field deflections. Transverse electron confinement is critical for realizing this concept experimentally. Using analytical theory, we show that the phase velocity of the laser pulse has a profound impact on the maximum transverse size of electron trajectories. The transverse size remains constant only below a threshold energy that depends on the degree of the superluminosity and it increases with the electron energy above the threshold. We illustrate this finding using 3D particle-in-cell simulations. The described increase can cause electron losses in tightly focused laser pulses, so it should be taken into account when designing high-intensity experiments at  high-power laser facilities.
\end{abstract}

\maketitle


\section{Introduction} 

Direct laser acceleration is one of fundamental mechanisms for transferring energy from an intense laser pulse to electrons of a laser-irradiated plasma. At relativistic laser intensities, most of the transferred energy is associated with the forward rather than transverse motion. This aspect has been successfully exploited for the development of
secondary particle (ion~\cite{Daido_2012,Macchi_2013_RMP}, neutron~\cite{higginson2011production,Pomerantz-PhysRevLett.113.184801}, positron\cite{cowan_positron,Chen_2009_PRL,Chen_2010_PRL}) and radiation sources~\cite{Schwoerer_ph,huang2016characteristics,Stark2016PRL} that have multiple inter-disciplinary applications.

There are two important aspects that distinguish direct laser acceleration in a plasma from that in a vacuum. In a vacuum, a laser beam of a finite width expels electrons radailly during the acceleration process. This limits the acceleration time for a given electron. Increased transverse velocity also increases electron dephasing from the laser pulse, which negatively impacts the energy gain. As the laser propagates through a plasma, it generates quasi-static radial electric and azimuthal magnetic fields~\cite{arefiev2016beyond}. These fields can provide electron confinement within the laser beam, leading to a significant increase in electron energy~\cite{pukhov1999DLA}. The energy gain is further facilitated by transverse electron deflections that alter electron dephasing~\cite{Khudik-POP_2016} (i.e. the slippage of the electron with respect to the laser wave-fronts).

The energy enhancement in the presence of the quasi-static plasma electric and magnetic fields is a threshold process~\cite{Arefiev_2012_PRL}. It has been previously established that the threshold depends on laser amplitude~\cite{arefiev2016beyond}, plasma field strength, and electron momentum at the start of the acceleration process (e.g. see Refs.~\onlinecite{Arefiev_2012_PRL, arefiev2016beyond,arefiev_JPP_2015,Khudik-POP_2016}). In experimentally relevant configurations, the width of the laser pulse is another important parameter that must be taken into consideration. 

In this paper, we examine the conditions for the  transverse electron confinement and show that the confinement depends on $(v_{ph} - c)/c$, where $v_{ph}$ is the phase velocity and $c$ is the speed of light. This parameter characterizes the relative degree of superluminosity. Our theoretical analysis shows that the maximum amplitude of transverse electron displacements is limited during the energy gain process only while the energy remains below a threshold value. Once the energy exceeds the threshold value, the transverse displacements begin to grow with energy. This increase in displacement can lead to electron losses and premature termination of the energy gain process in a tightly focused laser pulse. We provide multiple 3D particle-in-cell (PIC) simulations illustrating the described trajectory widening. 


\section{3D PIC simulation of trajectory widening} \label{Sec-widening}

In order to motivate the analytical analysis of Sec.~\ref{Sec-model}, we start with a 3D PIC simulation for a 700~TW laser pulse irradiating a uniform plastic target, initialized as a plasma consisting of fully ionized carbon ions and electrons. The laser peak intensity is $8.8 \times 10^{21}$~W/cm$^2$. The initial electron density is $n_e = 2n_{cr}$, where $n_{cr} \equiv m_e\omega^2/(4\pi e^2)$ is the classical critical density, $m_e$ and $e$ are the electron mass and charge, and $\omega$ is the frequency of the laser pulse. Table~\ref{table_PIC} provides additional information regarding the laser pulse, the target, and the simulation setup. {This and all other 3D simulations in the paper were performed using EPOCH~\cite{Epoch}. The photon emission module is turned off because the electron recoil is weak at the considered intensities.}

\begin{figure}
    \begin{center}
    \includegraphics[width=0.98\columnwidth]{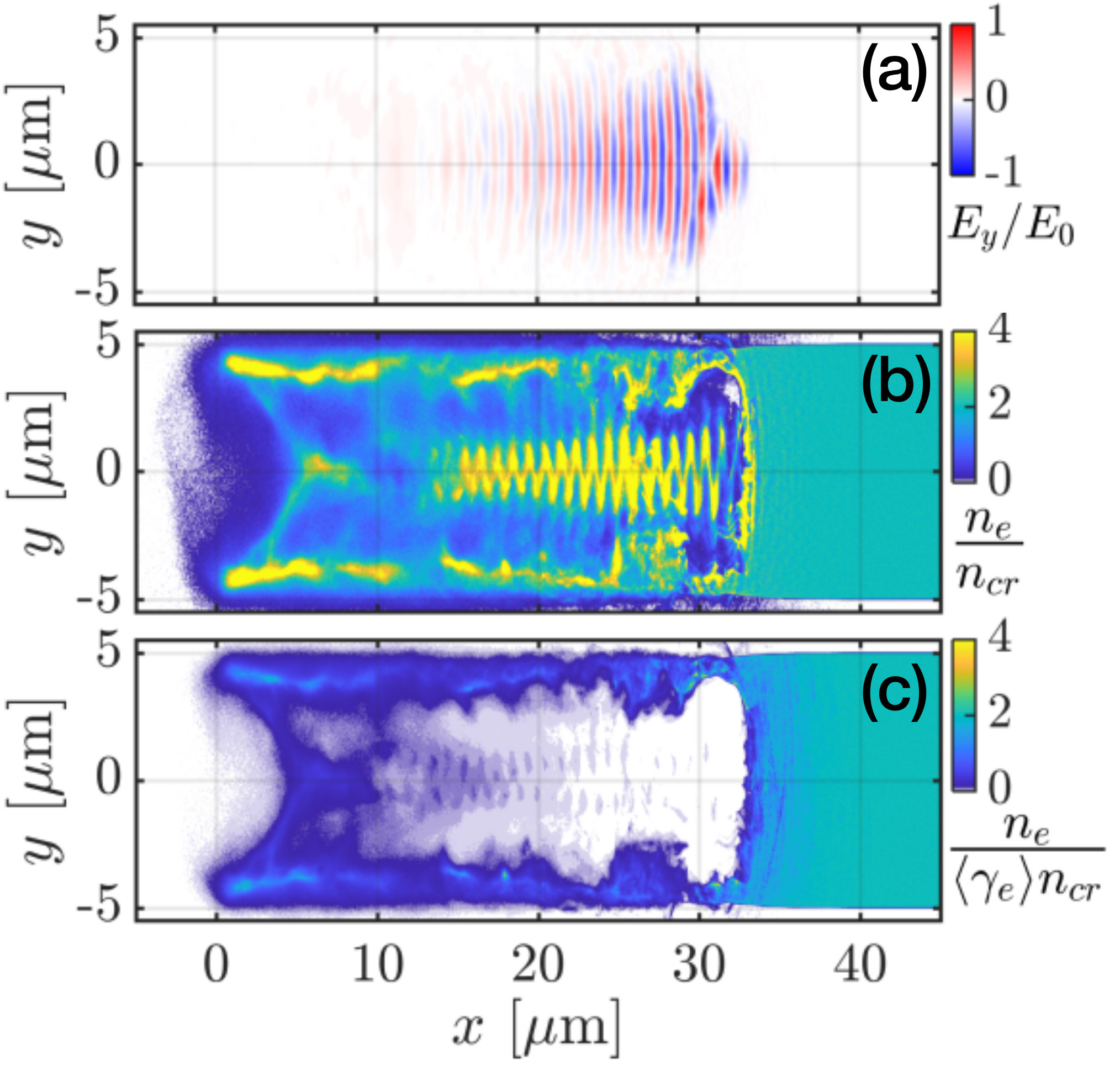}
    \caption{\label{fig:density} Laser propagation of the 700~TW laser beam due to relativistic transparency: (a) transverse electric field, (b) electron density normalized to $n_{cr}$, and (c) electron density normalized to  $\langle \gamma_e \rangle n_{cr}$, where $\langle \gamma_e \rangle$ is a cell-averaged value of the electron $\gamma$-factor. The snapshots are taken at $t = 104$ fs.}
    \end{center}
\end{figure}

Even though the electron density is higher than the classical critical density, the laser pulse is able to easily propagate through the target due to the relativistically induced transparency~\cite{gibbon2004shortRT,palaniyappan2012dynamics,fernandez2017laser,PhysRevLett.115.025002} caused by electron heating to relativistic energies. The transparency condition, $n_e \ll \langle \gamma_e \rangle n_{cr}$, is determined by the characteristic value of the electron $\gamma$-factor, $\langle \gamma_e \rangle$~\cite{gibbon2004shortRT}. Since the electrons are heated by the laser pulse, we estimate that $\langle \gamma_e \rangle \approx a_0$, with $a_0$ being the normalized laser amplitude defined as $a_0 \equiv |e| E_0 / (m_e c \omega)$, where $E_0$ is the peak amplitude of the laser electric field. The transparency condition then reads $n_e \ll a_0 n_{cr}$. It is well satisfied for the considered density $n_e = 2 n_{cr}$, since we have $a_0 = 80$ for the considered peak intensity. Figures~\ref{fig:density}(a) and  \ref{fig:density}(b) confirm that the laser pulse indeed propagates through the plasma rather than through an empty channel devoid of electrons. Figure~\ref{fig:density}(c) shows the electron density normalized to $\langle \gamma_e \rangle n_{cr}$ to illustrate the impact of the electron heating, where $\langle \gamma_e \rangle$ is the cell-averaged value of the electron $\gamma$-factor. Note that we define $t=0$~fs as the time when the laser pulse reaches its peak intensity in the focal plane in the absence of the target.



\begin{table}
\caption{Parameters used in the 3D PIC simulations}
\label{table_PIC}
\begin{tabular}{ |l|l| }
  \hline
  \multicolumn{2}{|c|}{3D PIC simulation parameters} \\
  \hline
  \multicolumn{2}{|l|}{\underline{\bf Laser pulse:} }\\
  Pulse energy & 6.5 and 26 J\\
  Peak intensity & $8.8 \times 10^{21}$ W/cm$^2$ \\
  $a_0$ & 80\\
  Polarization & linearly along $\mathbf{\hat{y}}$\\
  Wavelength & $\lambda$ = 1 $\mu$m \\
  Power & $P$ = 175 and 700 TW\\
  Location of the focal plane & $x$ = 0 $\mu$m, surface of plasma\\
  Pulse profile &\\
  (transverse \& longitudinal) & Gaussian \\
  Pulse duration & \\
  (FHWM for intensity) & 35 fs\\
  Pulse width/focal spot  &  \\
  (FWHM for intensity) & $w_0$ = 1.3 and 2.7 $\mu$m\\
  \multicolumn{2}{|c|}{}\\
  \multicolumn{2}{|l|}{\underline{\bf Plasma:} }\\
  Composition & carbon ions and electrons \\
  Target thickness &\\ (along $y$ and $z$) & $d=$ 10.0 $\mu$m\\
  Electron density & $n_e$ = 2 $n_{cr}$ \\
  Ion mass & $12m_p$ ($m_p$, mass of proton) \\
  Charge of ion & +6 $|e|$ \\
  Ion mobility & mobile \\
  Target length (along $x$) & $L$ = 70 $\mu$m \\
  \multicolumn{2}{|c|}{}\\
  \multicolumn{2}{|l|}{\underline{\bf General parameters:} }\\
  Spatial resolution & $30/\mu$m $\times 30/\mu$m $\times 30/\mu$m \\
  $\#$ of macro-particles/cell  &  \\
  Electrons & 15 \\
  Carbon ions & 5 \\
  \hline
\end{tabular}
\end{table}

\begin{figure}
    \begin{center}
    \includegraphics[width=1\columnwidth]{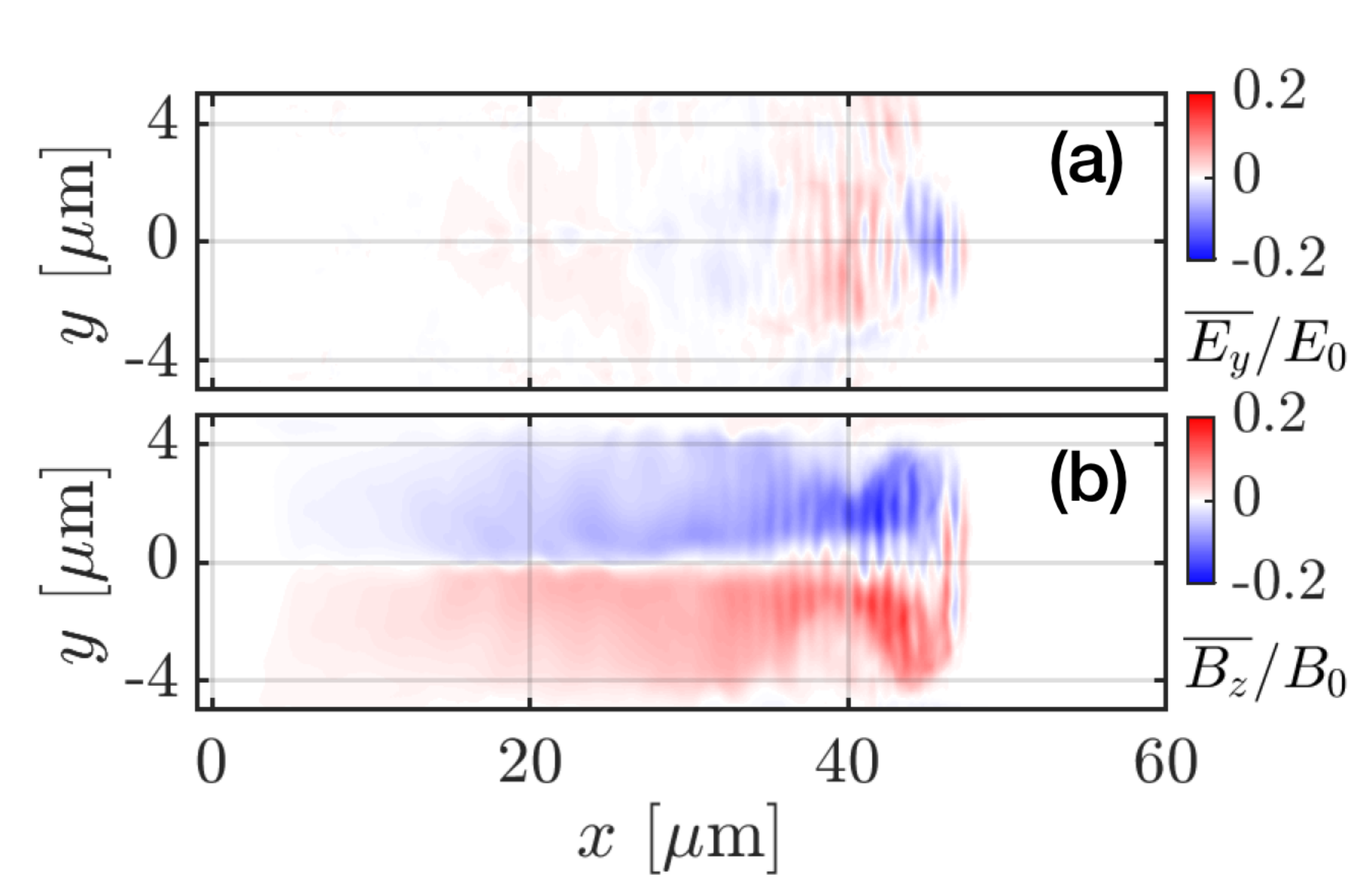}
    \caption{\label{fig:magnetic fields} Time-averaged transverse electric and azimuthal magnetic fields from the 3D PIC simulation for the 700 TW pulse. The fields are plotted in the $(x,y)$-plane located at $z=0$. $\overline{E_y}$ and $\overline{B_z}$ are the fields averaged over two laser periods. Both snapshots are taken when the electrons reach the highest cutoff energy, $\varepsilon_{max} \approx 750$~MeV. }
    \end{center}
\end{figure}

As the laser pulse propagates through the plasma, it generates a slowly evolving azimuthal magnetic field. The field is coiled around the axis of the laser beam, so we refer to it as a magnetic filament. The corresponding field structure is shown in Fig.~\ref{fig:magnetic fields}(b), where we plotted time-averaged $B_z$ in the $(x,y)$-plane located at $z=0$. Such a field facilitates the energy gain by laser-accelerated electrons via transverse deflections of these electrons. The laser can also generate a transverse quasi-static electric field by expelling plasma electrons that has a similar effect on the electron acceleration in the resulting channel~\cite{arefiev2016beyond}. However, the impact of this field compared to the effect of the magnetic field typically diminishes with laser amplitude. In our case, the quasi-static electric field shown in Fig.~\ref{fig:magnetic fields}(a) is weaker than the quasi-static magnetic field shown in Fig.~\ref{fig:magnetic fields}(b). Both fields are normalized to the peak laser electric and magnetic fields ($E_0$ and $B_0$) in vacuum in the absence of the target.

\begin{figure}
    \begin{center}
    \includegraphics[width=1\columnwidth,trim={0.5cm 0cm 1cm 0cm},clip]{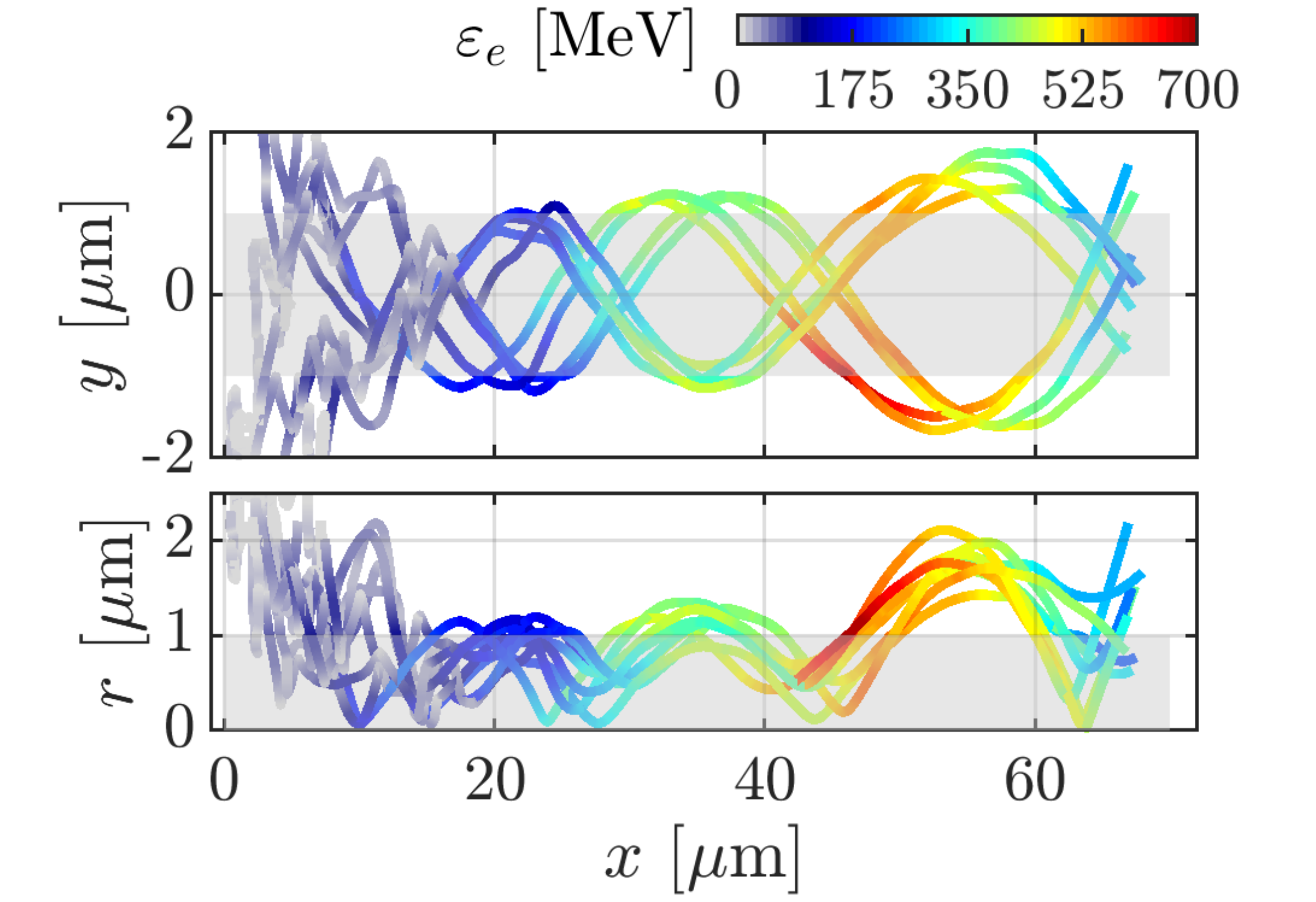}
    \caption{\label{fig:a0_80_traj} Electron trajectories in the 3D PIC simulation for the 700~TW laser beam. See Table~\ref{table_PIC} for more details.}
    \end{center}
\end{figure}

We have tracked all of the electrons in the simulation and we have randomly chosen 8 electrons with energies in the range between $(2/3)\varepsilon_{max}$ and  $\varepsilon_{max}$, where $\varepsilon_{max} \approx $~750 MeV is the maximum electron energy reached during the run. Figure~\ref{fig:a0_80_traj} shows the trajectories of the tracked electrons and their energy (color-coded). The upper panel is the projection onto the plane of the laser electric field polarization. The tracking confirms the described energy gain mechanism, with the electron energy increasing over multiple bounces across the magnetic filament as the electrons move in the positive direction along the $x$-axis.

Figure~\ref{fig:a0_80_traj} also shows that the amplitude of the transverse oscillations increases with electron energy. The only exception here is the early part of the electron trajectory with $x < 20$~$\mu$m that represents electron injection through the opening of the magnetic filament. It must be pointed out that the electrons develop strong oscillations not only in the plane of the laser electric field, but also out of this plane. The emergence of these non-planar orbits is likely related to the modulations of the electron $\gamma$-factor~\cite{ArefievPOP2016non-planar}. We plot the trajectories in the $(x,r)$-space in order to objectively assess their widening, where $r = \sqrt{y^2 + z^2}$ is the radius in the plane transverse to the axis of the laser beam. It is evident from the lower panel of Fig.~\ref{fig:a0_80_traj} that the maximum radius achieved by the electrons increases with each transverse bounce and that this increase is correlated with the energy increase.



\section{Test electron model} \label{Sec-model} 

Our goal is to identify the factors that limit the amplitude of the transverse electron oscillations inside the magnetic filament. In what follows, we call this amplitude the magnetic boundary. Motivated by the presented simulation, we consider a reduced model where a test electron is subjected to prescribed time-dependent fields of the laser and static fields of the channel. We neglect the electron recoil associated with photon emissions, i.e. the radiation-reaction force. Our approach to finding the magnetic boundary is to assume that the laser beam and the magnetic filament are wider than the magnetic boundary. We therefore approximate the laser by a plane electromagnetic wave with a superluminal phase velocity $v_{ph} > c$. The phase velocity in such a model can be used to account for the presence of the plasma and for the fact that the laser beam has a finite width~\cite{Robinson_PoP_2015}. In order to simplify our analytical derivations, we assume that the magnetic filament is sustained by a longitudinal current with a uniform current density $j_0$ (see Fig.~\ref{fig:magnetic fields lineout} in Sec.~\ref{Sec-power_scan} for supporting evidence).

The electron dynamics is described by the following equations:
\begin{eqnarray}
&& \frac{d \bm{p}}{d t} = - |e| \bm{E} - \frac{|e|}{\gamma m_e c} \left[ \bm{p} \times \bm{B} \right], \label{EQ_1} \\
&& \frac{d x}{d t} = \frac{c}{\gamma} \frac{p_x}{m_e c}, \label{EQ_2} \\
&& \frac{d y}{d t} = \frac{c}{\gamma} \frac{p_y}{m_e c}, \label{EQ_2_2} \\
&& \frac{d z}{d t} = \frac{c}{\gamma} \frac{p_z}{m_e c}, \label{EQ_2_3} 
\end{eqnarray}
where the electric and magnetic fields ($\bm{E}$ and $\bm{B}$) are given. Here $\gamma = \sqrt{1 + p^2 /m_e^2 c^2}$ is the relativistic $\gamma$-factor, $x$, $y$, and  $z$ are the electron coordinates, $\bm{p}$ is the electron momentum, and $t$ is the time. In the regime under consideration, $\bm{E} = \bm{E}_{wave}$ is just the laser electric field, whereas $\bm{B} = \bm{B}_{wave} + \bm{B}_{filament}$ is a superposition of the magnetic fields of the wave and the filament. Without any loss of generality, we consider a linearly polarized wave propagating in the positive direction along the $x$-axis with
\begin{eqnarray}
&& \bm{E}_{wave} = \bm{e}_y E_0 \cos (\xi), \label{E_wave} \\
&& \bm{B}_{wave} = \bm{e}_z \frac{c}{v_{ph}} E_0 \cos (\xi), \label{B_wave}
\end{eqnarray}
where $E_0$ is the wave amplitude and 
\begin{equation} \label{EQ-1.6}
\xi = \omega_0 t - \omega_0 x/ v_{ph}
\end{equation}
is the phase variable. The magnetic field of the channel sustained by the current density $\bm{j} = j_0 \bm{e}_x$ is given by
\begin{equation} \label{EQ-1.7}
\bm{B}_{filament} = \frac{m_e c^2}{|e|} \nabla \times \bm{a}_{filament},
\end{equation}
where
\begin{eqnarray} \label{EQ-8}
&& \bm{a}_{filament} = \bm{e}_x \frac{\alpha \left( y^2 + z^2 \right)}{\lambda_0^2} = \bm{e}_x \frac{\alpha r^2}{\lambda_0^2}, \\ \label{alpha_definition}
&& \alpha \equiv - \pi \lambda_0^2 |e| j_0 / m_e c^3,
\end{eqnarray}
and $\lambda_0 \equiv 2 \pi c / \omega_0$ is the vacuum wavelength. The filament that confines electrons has $j_0 < 0$. 

It can be verified using the equations of motion that the following quantity remains conserved as the electron moves in the considered field configuration:
\begin{equation} \label{C1}
     \gamma - \frac{v_{ph}}{c} \frac{p_x}{m_e c} + \frac{v_{ph}}{c} a_{filament} = C.
\end{equation}
We are going to consider a relativistic electron that is starting its motion on axis while moving in transverse direction. We specifically set 
\begin{eqnarray}
    && p_y = p_i, \\
    && p_z = 0
\end{eqnarray}
at $\xi = 0$ to mimic the electron injection into the magnetic filament observed in kinetic simulations~\cite{Stark2016PRL}. The constant of motion for this electron is its initial $\gamma$-factor $\gamma_i$:
\begin{equation}
    C = \gamma_i \equiv \sqrt{1 + p_i^2/m_e^2 c^2}.
\end{equation}

We first examine the electron dynamics in the absence of the laser field. In this case, the total momentum of the electron is conserved, which means that $C = \gamma$. We then find from Eq.~(\ref{C1}) that
\begin{equation} \label{Eq-19}
     \frac{p_x}{m_e c} = a_{filament}.
\end{equation}
There are two features of the electron dynamics that are evident from this relation. If $j_0 < 0$ and $a_{filament} \geq 0$, then the electron slides forward along the filament instead of performing rotations around a fixed axial location, since $p_x \geq 0$. The transverse displacements are constrained by the magnetic field,
\begin{equation} \label{r_MB_B=0}
    r \leq \lambda_0 \left( \frac{p_i}{\alpha m_e c} \right)^{1/2} \approx \lambda_0 \sqrt{\gamma_i / \alpha},
\end{equation}
with the corresponding limit obtained from Eq.~(\ref{Eq-19}) by setting $p_x = p_i$. The right-hand side is obtained by assuming that the electron is ultra-relativistic, $\gamma_i \approx p_i / m_e c$.

In the presence of the laser, the electron can gain energy from the laser field. In order to find how this impacts the radial confinement, we re-write Eq.~(\ref{C1}) as 
\begin{equation} \label{C1-v2}
     u a_{filament} = \gamma_i - \left[ \gamma - \frac{p_x}{m_e c} \right] + (u-1) \frac{p_x}{m_e c}.
\end{equation}
where we introduced
\begin{equation}
    u \equiv v_{ph} /c
\end{equation}
for compactness. As the longitudinal momentum and the $\gamma$-factor increase, the condition $\gamma -  p_x/m_e c > 0$ must hold. Moreover, we have $u - 1 \geq 0$. The maximum transverse displacement is achieved in the limit of $\gamma - p_x/m_e c \rightarrow 0$. In this limit, we can replace $p_x/m_e c$ with $\gamma$ in the last term, which yields
\begin{equation} \label{r_MB-main}
    r \leq r_\text{MB} \equiv \frac{\lambda_0}{\sqrt{\alpha u}} \left[ \gamma_i + (u - 1) \gamma  \right]^{1/2} .
\end{equation}
The right-hand side defines the radial location, $r = r_\text{MB}$, of the magnetic boundary. The trajectory remains confined within the magnetic boundary as the electron gains energy from the laser pulse.

We characterize the electron energy gain by the ratio $\gamma / \gamma_i$ and define two limiting cases based on the value of this parameter. If 
\begin{equation} \label{main-condition}
    \gamma/\gamma_i \ll (u-1)^{-1} = c / (v_{ph} - c),    
\end{equation}
then we have
\begin{equation} \label{MB-low}
     r_\text{MB} \approx \lambda_0 \sqrt{\gamma_i / \alpha u}.
\end{equation}
By comparing this expression with the one given by Eq.~(\ref{r_MB_B=0}), we conclude that in the presence of the laser the maximum radial displacement can become reduced due to the superluminosity ($u = v_{ph}/c > 1$).

In the opposite regime of 
\begin{equation} \label{main-condition-2}
    \gamma/\gamma_i \gg (u-1)^{-1} = c / (v_{ph} - c),    
\end{equation}
we have
\begin{equation} \label{MB-hight}
    r_\text{MB} \approx \lambda_0 \sqrt{\frac{u-1}{\alpha u}}\gamma^{1/2}.
\end{equation}
In this regime, the magnetic boundary expands as the electron energy increases, $r_{MB} \propto \gamma^{1/2}$.
On the other hand, the location of the magnetic boundary remains constant for $\gamma \ll \gamma_i / (u-1)$. Then the key conclusion is that there exists an electron energy threshold, approximately given by the condition 
\begin{equation}
    \gamma = \gamma_* \approx \gamma_i / (u-1),
\end{equation}
with the magnetic boundary becoming energy dependent above the threshold.


We have so far assumed that the magnetic field of the filament is much stronger than the radial quasistatic plasma electric field that arises due to charge separation. Such a regime has been observed in PIC simulations at $a_0 \gg 1$ and $n_e \gg n_{cr}$~[\onlinecite{Jansen_2018}]. At lower $a_0$ and lower plasma densities, the electric field can become comparable to or even stronger than the magnetic field~\cite{arefiev2016beyond}. Our results can be easily generalized to such a regime. We assume that the radial electric field is generated by a uniform charge density $\rho$. It can be verified using the equations of motion that include this field that the following quantity remains conserved:
\begin{eqnarray} \label{C2}
     && \gamma - \frac{v_{ph}}{c} \frac{p_x}{m_e c} + \frac{v_{ph}}{c} a_{filament} + \frac{\omega_{p0}^2 r^2}{4 c^2} \frac{\rho}{|e|n_0} = C,
\end{eqnarray}
where $n_0$ is the original electron density in the plasma and $\omega_{p0}^2 = 4 \pi n_0 e^2 / m_e$ is the corresponding electron plasma frequency. We now take into account Eq.~(\ref{EQ-8}) to obtain
\begin{eqnarray} \label{C3}
     && \gamma - \frac{v_{ph}}{c} \frac{p_x}{m_e c} + \kappa \alpha u  \frac{r^2}{\lambda_0^2} = C,
\end{eqnarray}
where
\begin{equation}
    \kappa \equiv 1  - \rho c^2/v_{ph} j_0 .
\end{equation}
In the limit of $\rho \rightarrow 0$, $\kappa = 1$ and Eq.~(\ref{C3}) reduces to Eq~(\ref{C1}).

In order to generalize the results given by Eqs.~(\ref{r_MB-main}), (\ref{MB-low}), and (\ref{MB-hight}), all we need to do is replace $\alpha$ by $\kappa \alpha$, where $\kappa$ must be calculated based on the current and charge densities in the filament. The magnetic field plays a major role in determining the electron dynamics if $\kappa \approx 1$, which is equivalent to the condition
\begin{equation}
    |\rho| c^2 \ll v_{ph} |j_0|.
\end{equation}
If $\rho > 0$ and $j_0 < 0$, then both the electric and magnetic fields confine the electron. This causes the radius of the magnetic boundary to become smaller, but it is important to stress that the threshold for the electron energy, $\gamma \approx \gamma_i / (u-1)$, remains unchanged. 


\section{Power scan} \label{Sec-power_scan}

In a laser beam with a given spot size, the widening of the magnetic boundary can lead to transverse electron losses. Therefore, the transverse electron confinement can be improved by widening the laser beam. In this section, we show that this approach makes it possible to achieve higher electron energies by increasing the laser power while keeping the same peak laser intensity.

\begin{figure}
    \begin{center}
    \includegraphics[width=0.98\columnwidth]{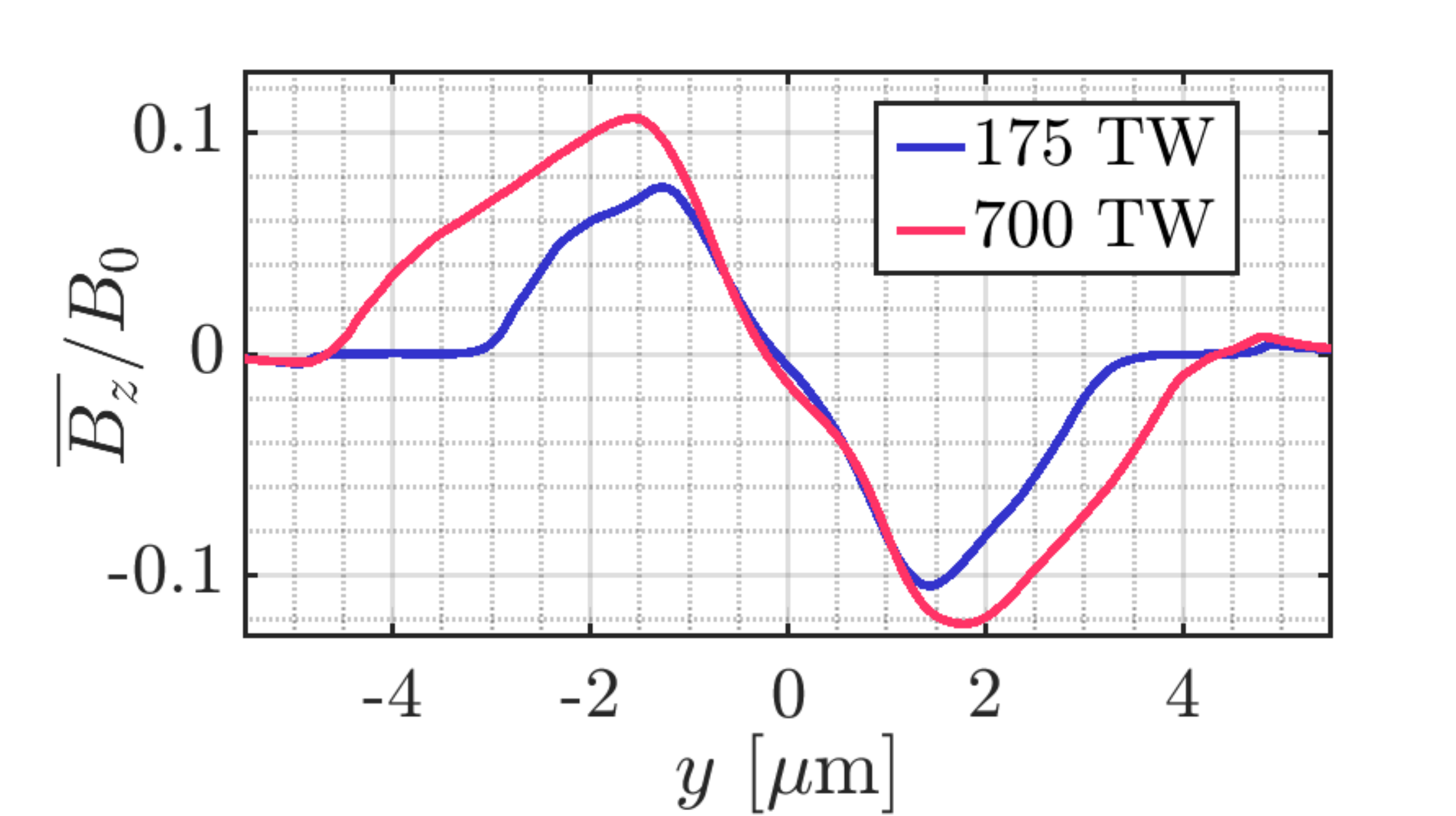}
    \caption{\label{fig:magnetic fields lineout} Transverse profiles of the quasi-static magnetic fields generated by the considered 175 TW and 700 TW laser pulses. For 175 TW (Fig.~\ref{fig:magnetic fields 200TW}), the profile is averaged over the region with 30 $\mu$m $\leq x \leq 34$~$\mu$m. For 700 TW (Fig.~\ref{fig:magnetic fields}), the profile is averaged over the region with 40 $\mu$m $\leq x \leq 45$~$\mu$m.}
    \end{center}
\end{figure}

\begin{figure}
    \begin{center}
    \includegraphics[width=1\columnwidth]{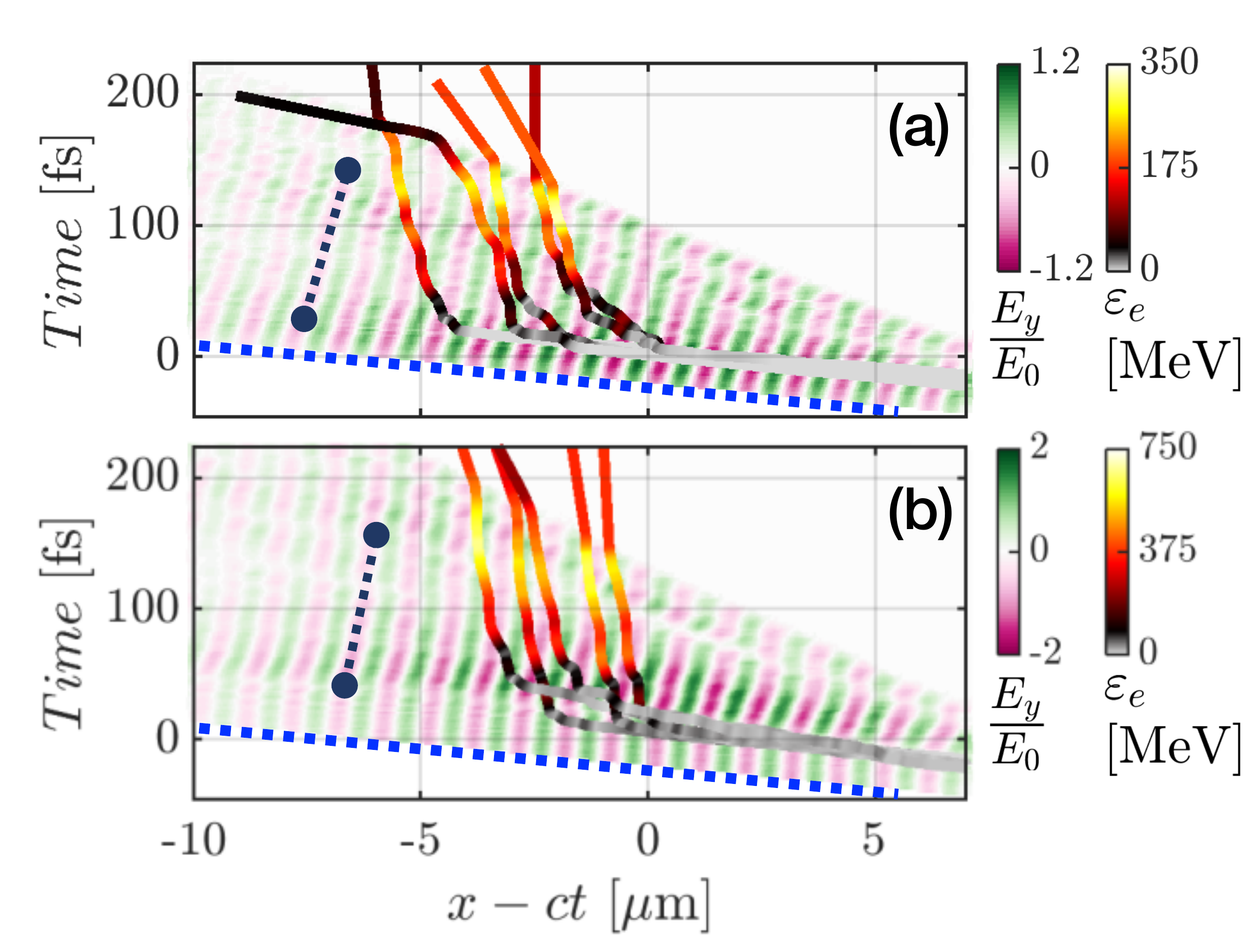}
    \caption{\label{fig:phase_velocity_traj} Temporal profiles of the laser electric field on the central axis ($y=z=0$) at (a) 175 TW and (b) 700 TW. The laser fields are plotted in a moving window together with longitudinal positions of randomly chosen energetic electrons. The black dashed lines show the segments used to find $v_{ph}$ in each run. The dashed blue line marks the left boundary of the simulation box located at $x = -5$~$\mu$m.}
    \end{center}
\end{figure}

We start by comparing our analytical results of Sec.~\ref{Sec-model} with the particle tracking shown in Sec.~\ref{Sec-widening}. Figure~\ref{fig:magnetic fields lineout} shows that the magnetic field varies linearly across the filament for $|y| < 1.5$~$\mu$m at $x = 0$~$\mu$m. This indicates that the current that sustains the magnetic field in this region can be approximated by a constant current density uniformly distributed in the cross-section. Based on the slope of the magnetic field, we find that this current density corresponds to {$\alpha \approx 17.3$} [see Eq.~(\ref{alpha_definition})]. In order to find the phase velocity, we have tracked the laser wave-fronts along the central axis [see {the black dashed line in} Fig.~\ref{fig:phase_velocity_traj}(b)]. In the region where the electron acceleration takes place, we have 
\begin{equation}
    u - 1 = (v_{ph} - c)/c \approx  2.5 \times 10^{-2}.
\end{equation}

\begin{figure}
    \begin{center}
    \includegraphics[width=1\columnwidth]{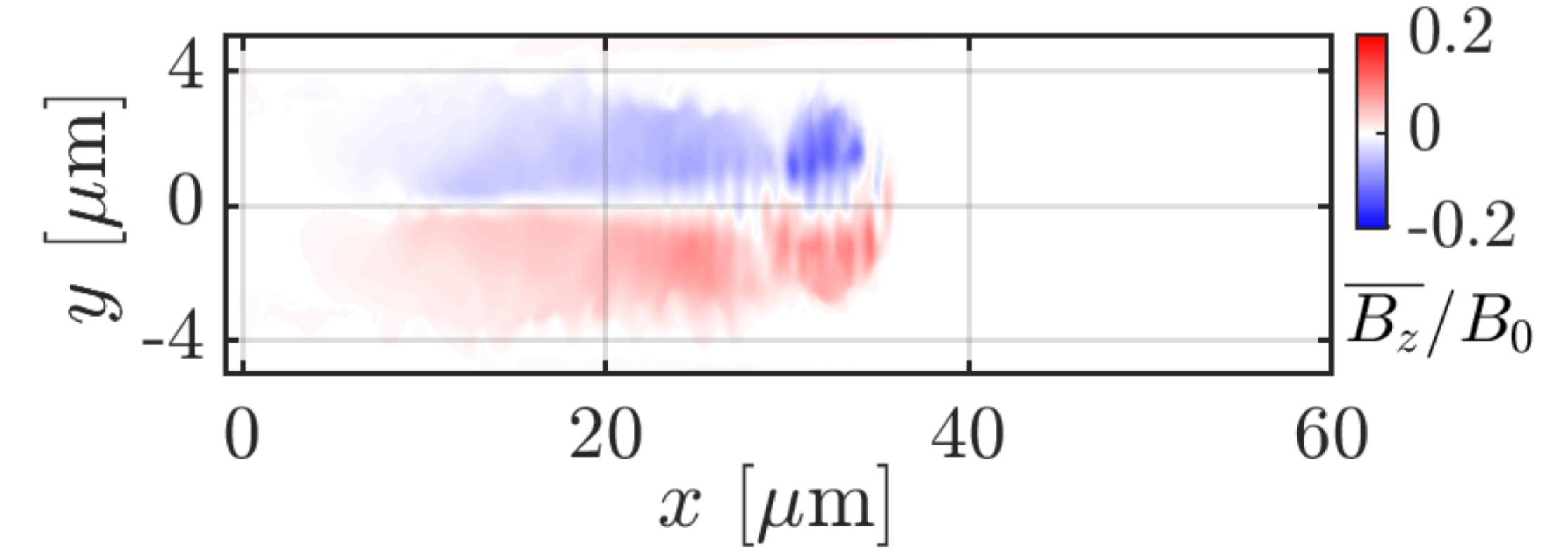}
    \caption{\label{fig:magnetic fields 200TW} Time-averaged azimuthal magnetic field from the 3D PIC simulation for the 175 TW laser pulse. The $z$-component of the magnetic field is plotted in the $(x,y)$-plane located at $z=0$. $\overline{B_z}$ is the field averaged over two laser periods. The snapshots are taken when the electrons reach the highest cutoff energy shown in Fig.~\ref{fig:spectrum}.}
    \end{center}
\end{figure}

As an example, we take one of the electron trajectories from Fig.~\ref{fig:a0_80_traj} and consider two subsequent radial deflections: the first one at about $x \approx 20$~$\mu$m and the second one at $x \approx 35$~$\mu$m. The maximum radius increases from 1.164~$\mu$m to 1.345~$\mu$m as the energy at the deflection point increases from 147~MeV to 411~MeV. According to Eq.~(\ref{r_MB-main}), these numbers correspond to $\gamma_i \approx 13.9$ if we offset the axis by 0.059~$\mu$m. The small offset (4.4\% of the radius) is a fitting parameter that reflects the fact that the axis of the filament is not perfectly aligned with the central axis ($x$-axis). {It is worth noting that the obtained $u$ and $\gamma_i$ correspond to a threshold energy of $\gamma_* m_e c^2 \approx 347$~MeV, which is consistent with the trajectories observed in Fig.~\ref{fig:a0_80_traj} where this energy is reached between the first and second deflection.} By applying Eq.~(\ref{r_MB-main}) to estimate the radius of the third deflection ($x \approx 55$~$\mu$m and $\varepsilon_e \approx 518$~MeV), we find that our theory under-predicts the transverse displacement. The actual deflection takes place at $r \approx 2.113$~$\mu$m. As seen from Fig.~\ref{fig:magnetic fields lineout}, this is outside of the region where the magnetic field changes linearly. This means that the electron travels radially through a region that has a lower magnetic field than what our  model assumes (the model assumes a linear slope) and this is likely the reason for the deflection point being further out radially.


We have performed another 3D simulation where we reduced the laser power by a factor of four to 175~TW while reducing the radius of the focal spot by a factor of two, such that the peak laser amplitude is the same as in the previous simulation. The details of the simulation are provided in Table~\ref{table_PIC}. A snapshot of the quasi-static magnetic field driven in this simulation is shown in Fig.~\ref{fig:magnetic fields 200TW}. The magnetic filament is more narrow than at 700~TW. However, as seen in Fig.~\ref{fig:magnetic fields lineout}, the slope of the magnetic field in the central region remains unchanged. This means that the current density in the central region also remains unchanged, with $\alpha \approx 17.3$. The phase velocity is slightly higher at lower power due to the beam being more narrow, with $u - 1 \approx  2.85 \times 10^{-2}$ (see Fig.~\ref{fig:phase_velocity_traj}).

\begin{figure}
    \begin{center}
    \includegraphics[width=1\columnwidth,trim={0.5cm 0cm 1cm 0cm},clip]{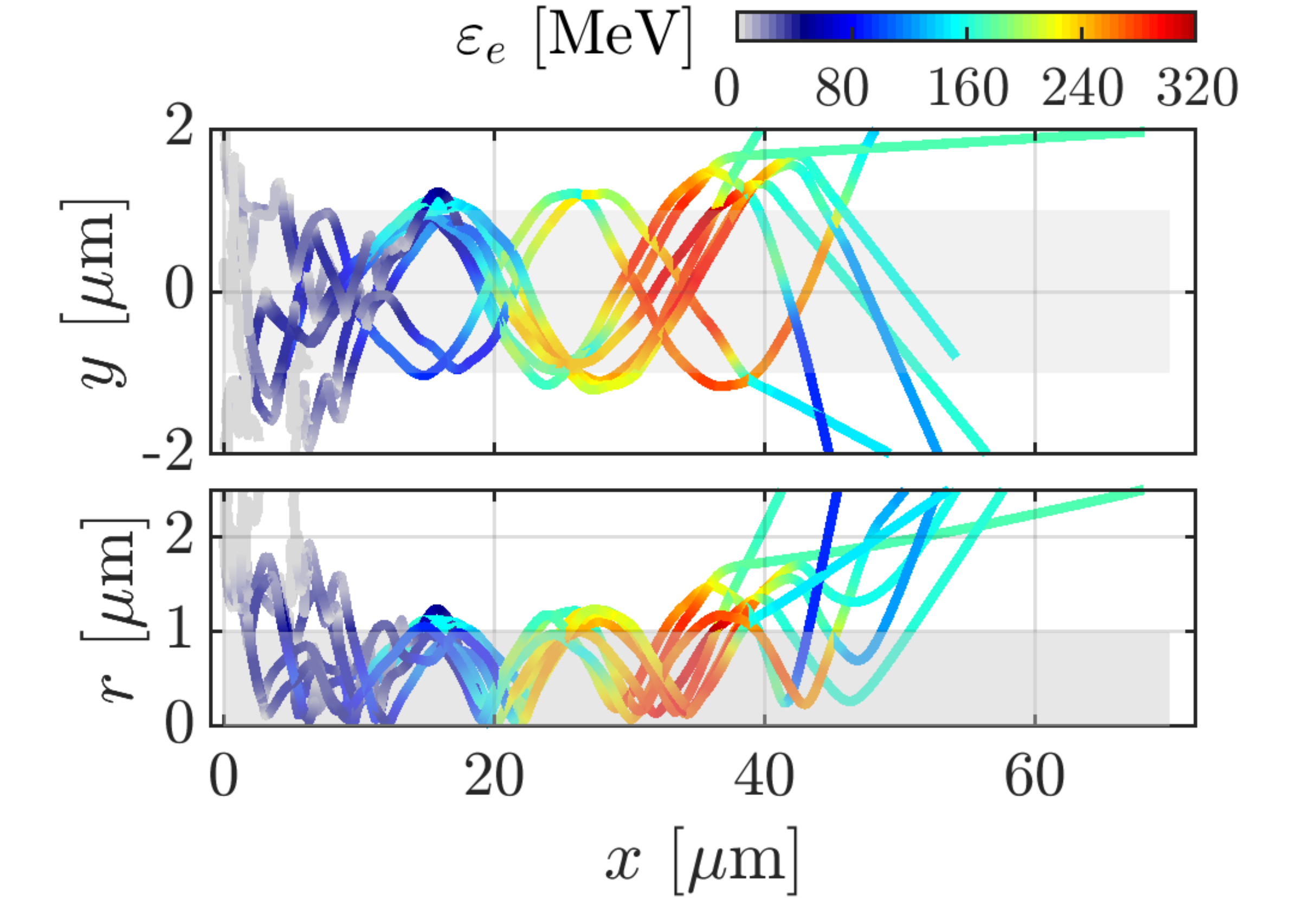}
    \caption{\label{fig:a0_80_traj_200TW} Electron trajectories in a 3D PIC simulation for a laser pulse with $a_0=80$ and power of 175 TW. See Table~\ref{table_PIC} for more details.}
    \end{center}
\end{figure}

Figure~\ref{fig:a0_80_traj_200TW} shows trajectories of energetic electrons at 175~TW, where we again randomly chose eight electrons with energies in the range between $(2/3)\varepsilon_{max}$ and $\varepsilon_{max}$. {Here $\varepsilon_{max} \approx 380$~MeV.} In contrast to the run for 700~TW, the electrons are lost at much lower energy. The losses are likely associated with the fact that the filament is more narrow, which limits the increase of the magnetic boundary with energy. Figures~\ref{fig:phase_velocity_traj}(a) and \ref{fig:phase_velocity_traj}(b) additionally show the longitudinal positions of randomly chosen energetic electrons in the two runs that are being compared (175 and 700~TW). The key feature of these plots is that they confirm that the electrons are indeed injected transversely into the laser beam. The injection location can be roughly identified by the first major bend along a given electron trajectory.

\begin{figure}
    \begin{center}
    \includegraphics[width=0.95\columnwidth]{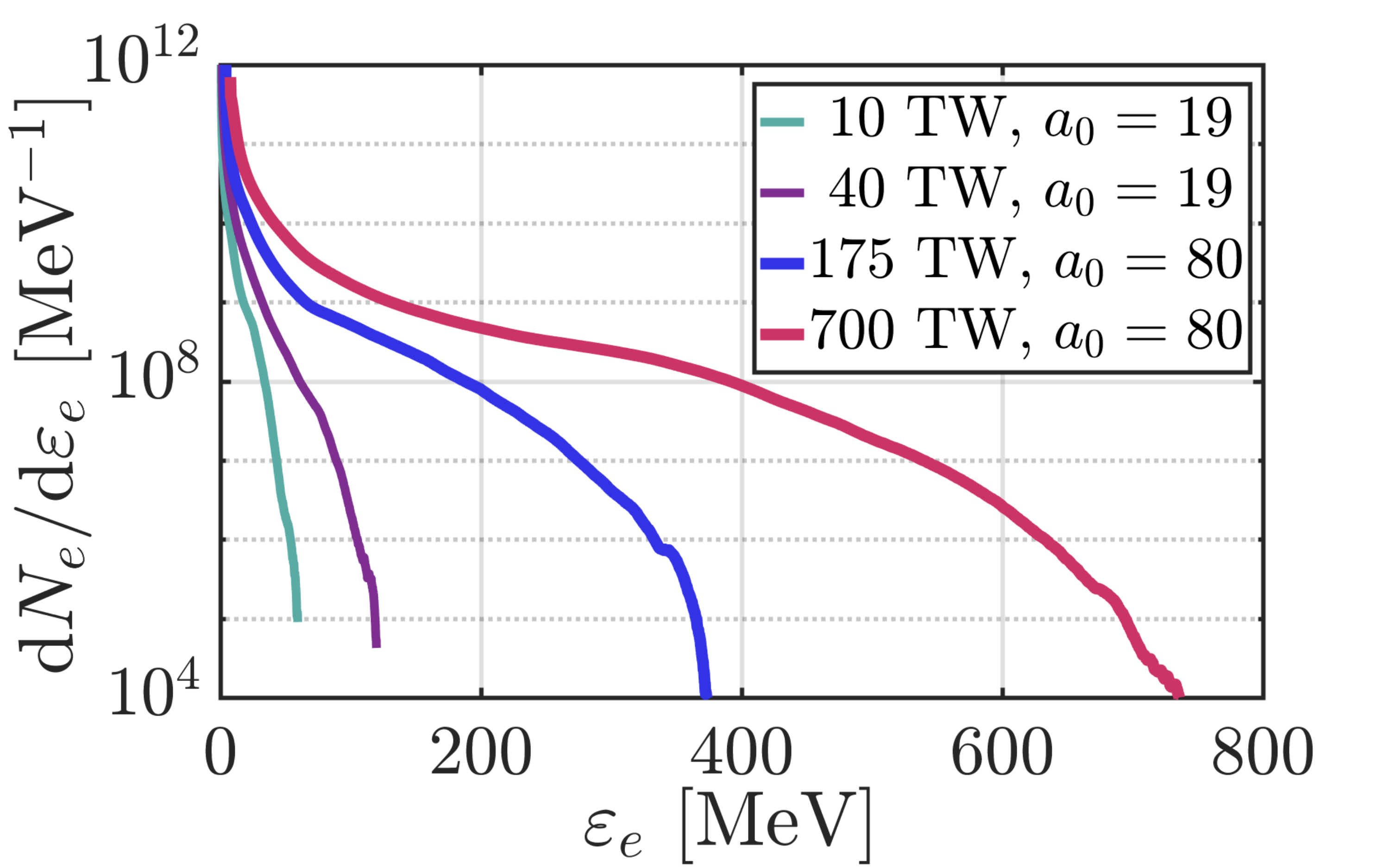}
    \caption{\label{fig:spectrum} Snapshots of electron spectra from 3D PIC simulations. The snapshots are taken when the electrons reach the highest cutoff energy, i.e. at $t = 116$~fs for 175~TW, $t = 162$~fs for 700~TW, $t = 72$~fs for 10~TW, and $t = 82$~fs for 40~TW. In each case, the spectra are calculate for the entire electron population in the simulation box. The ``temperature'' representing the slope of the energetic tail is 3.5~MeV for 10~TW, 6.0~MeV for 40~TW, 24~MeV for 175~TW, and 45~MeV for 700~TW.}
    \end{center}
\end{figure}

Figure~\ref{fig:spectrum} shows the electron spectra in the two considered simulations. The snapshots are taken when each of the spectra reach their maximum electron energy
(at $t = 116$~fs for the 175~TW run and at $t=162$~fs for the 700~TW run). It is evident that the laser-accelerated electrons are able to achieve much higher energy in the wider magnetic filament, i.e. at higher laser power. This observation holds at lower intensity of $a_0 = 19$. We again varied the power while keeping the peak intensity fixed (all other parameters are the same as in Table~\ref{table_PIC}, with the smaller $w_0$ corresponding to the 10~TW run). Even though the transverse quasi-static electric field is likely to play a more prominent role in these simulations, the observation that a wider filament with quasi-static confining fields allows for an increased energy gain holds.

\section{Summary}

We have shown that a laser-driven electron can gain energy without increasing the amplitude of its transverse oscillations only while $\gamma \ll \gamma_i c/(v_{ph} - c)$, where $\gamma_i$ is the initial electron energy. The condition essentially provides a threshold energy. The amplitude of the transverse oscillations that we call the magnetic boundary radius starts to increase with electron energy once the value of $\gamma$ exceeds the threshold value. The non-trivial conclusion is that the threshold is dependent on the superluminosity of the laser.

Using 3D PIC simulations, we found that the quasi-static magnetic field of the filament has two distinct regions: the central region where its amplitude increases linearly with radius and the outer region where a rollover occurs, leading to a gradual ramp down of the amplitude to zero. Our model describes well the electron dynamics in the central region. Once the magnetic boundary reaches the outer region due to the energy increase, the electrons can lose their confinement.


An important takeaway message from our analysis is that there could be a significant benefit from a power increase even at constant intensity~\cite{Tao_power_scan}: the electrons can gain higher energy in a wider beam before being lost due to the boundary expansion. In terms of a single laser system with a given total energy, our results provide an additional consideration when deciding how to focus the laser pulse. The benefits of increasing the peak intensity must be evaluated against the possibility that tighter focusing would lead to premature electron losses during the acceleration process due to a reduced beam width.

In this paper, we focused on the regimes where the electron recoil due to photon emissions by the energetic electrons can be neglected. In fact, we have verified that the spectra shown in Fig.~\ref{fig:spectrum} remain essentially unchanged if the recoil is included into the simulations. It has been previously shown for a laser pulse with $v_{ph} = c$ that the electron recoil causes a contraction of the magnetic boundary~\cite{gong_SR_2019}. Our results show that the superluminosity has an opposite effect on the magnetic boundary. Therefore, a regime where both effects are important is likely to lead to a non-trivial evolution of the magnetic boundary.

\section*{Acknowledgements}

This research was supported by AFOSR (Grant No. FA9550-17-1-0382). Simulations were performed with EPOCH (developed under UK EPSRC Grants No. EP/G054940/1, No. EP/G055165/1, and No. EP/G056803/1) using HPC resources provided by TACC at the University of Texas. This work used XSEDE, supported by NSF grant number ACI-1548562.

\section*{References}
\bibliography{Collection}


\end{document}